\pdfoutput=1

\documentclass[11pt]{article}

\usepackage[final]{acl}

\usepackage{times}
\usepackage{latexsym}
\usepackage{hyperref}
\usepackage{url}
\usepackage[capitalise]{cleveref}

\usepackage[T1]{fontenc}

\usepackage[utf8]{inputenc}

\usepackage{microtype}

\usepackage{inconsolata}

\usepackage{graphicx}

\usepackage[moderate]{savetrees}

\title{Policy Frameworks for Transparent Chain-of-Thought Reasoning\\ in Large Language Models}

\author{Yihang Chen, Haikang Deng, Kaiqiao Han, Qingyue Zhao\\
Department of Computer Science\\
University of California, Los Angeles\\
  \texttt{{\{yhangchen, haikang, kqhan, zhaoqy24\}}@cs.ucla.edu}\\
  }

\begin{document}

\maketitle
\begin{abstract}
Chain-of-Thought (CoT) reasoning enhances large language models (LLMs) by decomposing complex problems into step-by-step solutions, improving performance on reasoning tasks. However, current CoT disclosure policies vary widely across different models in frontend visibility, API access, and pricing strategies, lacking a unified policy framework. This paper analyzes the dual-edged implications of full CoT disclosure: while it empowers small-model distillation, fosters trust, and enables error diagnosis, it also risks violating intellectual property, enabling misuse, and incurring operational costs. We propose a tiered-access policy framework that balances transparency, accountability, and security by tailoring CoT availability to academic, business, and general users through ethical licensing, structured reasoning outputs, and cross-tier safeguards. By harmonizing accessibility with ethical and operational considerations, this framework aims to advance responsible AI deployment while mitigating risks of misuse or misinterpretation.
\end{abstract}

\section{Motivation}
\label{sec:motivation}
Chain-of-Thought (CoT) reasoning~\citep{wei2022chain} in large language models (LLMs) is a technique in which models are prompted to generate intermediate reasoning steps before arriving at a final answer. This method not only improves model performance on tasks that involve reasoning skills, such as multi-step mathematical problem-solving and logical inference~\citep{wei2022chain, sprague2024cotcotchainofthoughthelps} but also increases model transparency by exposing a trace of its decision-making process~\citep{wu2022aichainstransparentcontrollable}. In this paper, we are motivated toward developing a fine-grained policy framework to govern the level of access for LLM users.

The introduction of CoT brings a great shift in how LLMs are used to solve complex problems. In the past, researchers have been using LLMs for end-to-end question answering, where models generate responses based on patterns learned from training data and produce answers directly in their generation. While this approach leverages a model’s capability to recognize and replicate patterns, it does not fully utilize a model’s reasoning skills. In fact, models often struggle with complex tasks that require logical reasoning. In contrast, CoT decomposes a problem into sequential steps which require the model to explicitly structure its thought process, mimicking human problem-solving strategy. By doing so, CoT enhances a model's ability to tackle tasks that require logical deduction, arithmetic calculations, and contextual understanding across multiple sentences or paragraphs.


Moreover, one key advantage of CoT is its ability to enhance performance on reasoning-intensive tasks, as demonstrated by \citet{wei2022chain}, who show that CoT significantly improves LLM accuracy on benchmarks like GSM8K (Grade School Math). Similarly, \citet{kojima2023largelanguagemodelszeroshot} explore a zero-shot variation of CoT, revealing that LLMs can generate intermediate steps to guide them toward more reliable answers even without fine-tuning, underscoring CoT’s inherent value in LLMs. Furthermore, CoT serves as a foundation for more advanced and interpretable reasoning methods, such as self-consistency decoding introduced by \citet{wang2023selfconsistency}, which selects the most consistent answer across multiple attempts. Additionally, program-aided language models proposed by \citet{gao2023palprogramaided} integrate external tools to enhance the reliability of CoT-based inference, highlighting its crucial role in improving reasoning frameworks.

Another important aspect of CoT is its role in reducing hallucinations in LLM outputs. A major issue of LLM is its tendency to generate confident but factually incorrect content. \citet{Ji_2023} observes that by encouraging the model to follow a structured sequence of steps rather than producing answers directly, they can lessen LLM hallucinations in various kinds of tasks. It is the process of breaking down the task in CoT that reduces the likelihood of reasoning shortcuts that lead to incorrect conclusions.

Furthermore, CoT also offers advances in model interpretability and transparency. LLMs have long been referred to as ``black-boxes’’ due to their opacity and lack of explainability. This hinders the deployment of LLMs in high-stakes applications such as medical diagnosis, legal reasoning, and scientific analysis. CoT, however, allows human to inspect the model’s reasoning path, making it easier to diagnose the errors made during generation. \citet{wu2022aichainstransparentcontrollable} emphasizes that CoT-based reasoning helps relieve concerns about unreliable AI responses by enabling human users to analyze each step of the reasoning process. 



With CoT's trait of disclosing LLM reasoning paths and the properties mentioned above, it is crucial to gate users access to CoT outputs based on their identities. The necessity arises as different users require varying level of access to CoT outputs depending on their expertise, intent, and use case. A tiered access framework would render safer, more controlled use of CoT reasoning.

The contributions of this paper can be summarized as follows:  
\begin{itemize}  
    \item \textbf{Pioneering CoT Disclosure Framework}: We are the first to recognize the importance of CoT disclosure and propose a policy framework to regulate it. Additionally, we conduct the first survey on different models regarding frontend visibility, API access, and pricing strategies, highlighting the lack of a unified policy framework.  
    \item \textbf{Comprehensive Analysis of CoT Disclosure}: We analyze the benefits and risks of full CoT disclosure, emphasizing its role in small-model distillation, trust-building, and error diagnosis while addressing concerns related to intellectual property, misuse, and operational costs.  
    \item \textbf{Tiered-Access Policy Framework}: We propose a structured approach to CoT availability by categorizing users into academic, business, and general tiers, ensuring a balance between transparency, accountability, and security.  
\end{itemize}

\section{Transparency of CoT in Current LLMs}\label{sec:trans}

\paragraph{Frontend}

As of March, 2025, frontend interfaces of most recently released proprietary reasoning models like OpenAI o3-mini, DeepSeek R1, and Grok 3 \citep{o3mini2024,guo2025deepseek,grok3} provide CoTs rendered in ``semi''-rich-text formats to even free users.
OpenAI o1 \citep{jaech2024openai} originally did not disclose CoTs through their Web interface. After the release of DeepSeek-R1, whose frontend displays very detailed and seemingly raw CoTs, the OpenAI chat frontend began to output rendered CoTs (including LaTeX formats) to some extent on the fly.
According to the personal experience of the authors, when confronting the same math puzzle, the DeepSeek frontend often tends to display relatively thorough CoTs, which usually contain obvious pattern words of self-reflection like ``Wait, '' and ``But ...''; the OpenAI frontend (March, 2025), in contrast, chooses to hide the full CoTs by default and only display a decently short CoT even after the user try to check it, it sometimes only displays a succinct CoT that appears to be a summary of the reasoning process for users to understand the model's decision.
Actually, it is a common issue in the disclosed CoTs (from Web frontends of popular proprietary models) that they are sometimes too coarse-grained and give up diving into the problem details too early, but are sometimes very leng and appear to be superficial paraphrases of the problem in stitches without obvious key breakthroughs.
Patterns of CoTs like these are referred to as ``overthinking'' \citep{cuadron2025danger}.
\citet{cuadron2025danger} has made some preliminary attempts to quantify the quality of CoTs, especially through the lens of ``metric/score of overthinking'', where three identified patterns of overthinking and the associated scoring mechanism (through LLM-as-a-judge systems) shed light on whether and in which case the frontend should disclose certain parts (or summarizations/paraphrases) of CoTs to users.

\paragraph{API Pricing} Through APIs of OpenAI's flagship reasoning models including o1, o3, and o3-mini, the reasoning trajectory is entirely invisible to users through the API but tokens in these CoTs are charged as expensive as visible output tokens returned as answers \citep{openaiAPI}. DeepSeek employs the same pricing strategy for \verb|deepseek-reasoner| but return the CoTs content via their API \citep{deepseekAPI}.
The API of Claude 3.7 also gives an option of returning reasoning tokens, but only charges for tokens in the final answer \citep{claudeAPI}.
The rules of CoT disclosure (again, as of March, 2025) in these leading companies may have some downstream issues as follows.

\paragraph{Status Quo Implications}
On one hand, the inference cost of generating reasoning tokens for autoregressive LLMs is not significantly lower than that of finalized output tokens, which means the extra rate charged for CoT disclosure could be reasonable. On the other hand, keeping CoTs totally secret, as done via OpenAI's API, may not only leads to accountability issues but also implicitly let users become neglectful of the importance of the reasoning details, especially those API users outside of the AI research community. From this aspect, DeepSeek's current disclosure rule is plausibly a good compromise between the two extremes, and its status quo indeed has some positive implications. For example, fine-grained distillation of key reasoning steps \citep{dai2024beyond} become doable with the aid of full CoTs from powerful LLMs; moreover, since DeepSeek allows API users to set \verb|max_tokens| \emph{after} CoT to be $1$, users can nearly only obtain the CoT from the DeepSeek API and feed it to cheaper LLMs to get the final answer in the interest of financial cost.\footnote{\url{https://x.com/skirano/status/1881854481304047656}}
However, our understanding of CoTs, from both theoretical \citep{feng2023towards} and empirical \citep{dutta2024think} points of view, especially in terms of the sensible quality of CoTs \citep{cuadron2025danger}, is still quite elementary. As mentioned in the first paragraph of this section, CoTs with overthinking, which could be very brief, might be misleading for both customer and business users. Thus, the limited understanding implies that there is by far no affirmative conclusion on whether disclosing CoTs ``as is'' is beneficial from the perspective of users. The dilemmas outlined in this paragraph are further discussed in detail in Section~\ref{sec:pro} and Section~\ref{sec:con}, respectively.

\section{Arguments pro Transparent CoT}\label{sec:pro}
\subsection{CoT Empowers Small-sized LLM}
\label{sec:pro_cot_1}
Modern state-of-the-art reasoning models are characterized by immense scale. For instance, DeepSeek-R1~\citep{guo2025deepseek} is a 671B parameter mixture-of-experts (MoE) model with 37B activated parameters per inference. While the exact size of OpenAI’s o1~\citep{jaech2024openai} and o3-mini~\cite{o3mini2024} series remains undisclosed, industry experts speculate they operate on the order of 100B parameters~\citep{abacha2024medec}. Due to their computational and memory demands, such large models are impractical for on-device deployment.

For edge or device-side applications, smaller, dense architectures are more feasible. However, reinforcement learning (RL)—a common training method for large models—is less effective when applied to small-sized models~\citep{guo2025deepseek}. Instead, a promising approach involves distilling the CoTs of larger, more capable models into smaller ones. Empirical results show this strategy’s efficacy.  DeepSeek-R1-Distill-Qwen-7B outperforms prior open-source models on AIME 2024, MATH-500, and LiveCodeBench and rivaling o1-mini.

 Further supporting this paradigm, recent studies demonstrate that minimal high-quality reasoning traces can unlock significant gains. For example, fine-tuning the {Qwen2.5-32B-Instruct} model on {s1K}~\citep{muennighoff2025s1}—a dataset of 1,000 questions with curated reasoning traces—and augmenting it with budget forcing produces the model s1-32B, which exceeds o1-preview by up to 27\% on competition math benchmarks (MATH and AIME24). This aligns with the \textbf{L}ess-\textbf{I}s-\textbf{M}ore \textbf{R}easoning (\textbf{LIMO},~\cite{ye2025limo}) Hypothesis, which speculates that in foundation models with comprehensive pre-training, sophisticated reasoning can emerge through minimal but precisely structured demonstrations of cognitive processes.

These findings highlight the critical role of high-quality reasoning pattern demonstrations in enhancing small LLMs. Transparent disclosure of state-of-the-art models’ CoT provides a blueprint for distilling advanced logic. By leveraging these insights, on-device models can achieve performance once exclusive to large-scale systems, accelerating the integration of LLMs into daily life—from personalized education tools to real-time decision support. Besides, by enabling high-performance small models, CoT disclosure reduces reliance on proprietary, centralized systems (e.g., OpenAI’s undisclosed o1/o3-mini), fostering open innovation and equitable access to AI tools.

\subsection{CoT Builds Trust}
\label{sec:pro_cot_2}
When artificial intelligence increasingly shapes critical decisions, transparency in AI operations is essential. Evidence~\citep{zhou2025hidden,arrieta2025o3} shows that the stronger the model's reasoning ability, the greater the potential harm it may cause when answering unsafe questions.
Full CoT disclosure—the practice of revealing an AI’s step-by-step reasoning—offers profound benefits that enhance trust, accountability, and ethical integrity.

Transparency in AI’s reasoning fosters trust and accountability. When users see the logical steps behind an AI’s conclusions—such as a medical diagnosis system linking symptoms to conditions—they gain insight into its reliability, reducing the "black box" problem. This clarity makes AI as a collaborative partner rather than an incomprehensible oracle. Recently, MedVLM-R1~\citep{pan2025medvlmr1incentivizingmedicalreasoning} introduces reasoning into the medical image analysis to improve transparency and trustworthiness. 
Moreover, CoT disclosure enables precise error correction. For instance, if an autonomous vehicle causes a collision, tracing its reasoning could allows engineers to address the issue. This accountability ensures developers remain responsible for ethical oversight, driving iterative improvements and aligning AI evolution with societal needs.

Additionally, CoT disclosure mitigates bias and upholds ethical standards. AI systems risk perpetuating biases from training data, but exposing their logic allows scrutiny of discriminatory patterns. A hiring tool favoring specific demographics, for example, could be flagged and corrected if its CoT reveals biased criteria. Beyond technical fixes, transparency respects user autonomy by clarifying how decisions impacting lives are made—a cornerstone of ethical AI. These safeguards align with global demands for responsible innovation, ensuring AI operates as a force for equity rather than exclusion.

\section{Arguments against Transparent CoT}\label{sec:con}

\subsection{CoT Exposes Proprietary Advantages} 
The full disclosure of CoT reasoning traces could inadvertently expose LLM secrets. Companies such as OpenAI, Deepseek, and Anthropic have developed advanced reasoning techniques that distinguish their models from open-source alternatives. These methodologies, including unique prompting and training strategies and datasets, contribute to state-of-the-art performance. Full disclosure of CoT reasoning would effectively open these innovations to competitors.
Moreover, revealing reasoning traces can facilitate adversarial attacks, as attackers could analyze these traces to identify vulnerabilities or generate adversarial prompts that manipulate model behavior~\cite{su2024enhancingadversarialattackschain}. This risk is particularly concerning for safety-critical applications, where maintaining robustness against adversarial exploitation is paramount~\cite{xiang2024badchainbackdoorchainofthoughtprompting}.

\subsection{Risks of Misuse and Misinterpretation} \label{subsec:misuse}
Full CoT disclosure could lead to widespread misinterpretation and misuse. Many users lack the technical expertise to critically evaluate reasoning traces, potentially leading to overconfidence in AI decisions even when the reasoning is flawed~\citep{turpin2023languagemodelsdontsay}.
For example, detailed reasoning traces in medical AI applications might provide a false sense of security to practitioners lacking the domain expertise to detect subtle errors. A flawed logical step in an AI-generated medical diagnosis could mislead users into accepting incorrect conclusions, exacerbating risks rather than mitigating them.
Furthermore, bad actors could exploit CoT transparency to generate deceptive content. If CoT traces from highly capable models are available, they could be repurposed to enhance misinformation campaigns, fraud, or automated social engineering attacks. Malicious agents could manipulate users more effectively by mimicking the logical coherence of trustworthy AI-generated content~\citep{xiang2024badchainbackdoorchainofthoughtprompting,han2024conceptreversedwinogradschemachallenge}.

\subsection{Inherently Misleading CoT}\label{subsec:inherent}
Raw CoTs can still sometimes misguide users on low-stakes tasks including basic coding and math puzzles.
As outlined in Section~\ref{sec:trans}, given clearly executable coding instructions, the agent may generate excessive yet superficial planning without executing actions, or generate mutually dependent steps in a single turn before external feedbacks are demonstrated and thus form deadlocks, or abandon/conclude tasks based solely on internal reasoning rather than validating outcomes through environmental interaction \citep{cuadron2025danger}. Quantitative measurement or alleviation of these overthinking effects are relatively open, which undermines the accessibility of raw CoTs in general in the short term. Additionally, performative reasoning models sometimes switch among multiple languages in their CoTs\footnote{\url{https://x.com/RishabJainK/status/1877157192727466330}}, which might contain less understandable languages or even low-resource languages. This emergent phenomenon may have unintended consequences for users who only master one language when revealing CoTs even after an imperfect machine translation procedure (into the users' preferred language). Conceptually, the vendors can develop a set of ``philosophical language only for the purpose of reasoning''\footnote{\url{https://www.gilesthomas.com/2025/01/philosophical-language-llm}}, and decode this type of CoT into human-readable language before returning it to users; but this might be a long-standing open problem in the foreseeable future.

\subsection{Computational \& Operational Costs}
Generating, storing, and validating detailed CoT reasoning steps at scale introduces significant computational and operational overhead~\citep{chen2025unveilingkeyfactorsdistilling}. Unlike simple model outputs, CoT explanations require additional processing, increasing latency and resource consumption. For real-time applications, such as on-device assistants or financial trading systems, the additional burden of maintaining exhaustive CoT logs may render deployment impractical.
If full CoT disclosure becomes a standard expectation, AI companies may need to invest heavily in infrastructure to validate and refine reasoning traces.~\citep{cui2025stepwiseperplexityguidedrefinementefficient}.

\section{Policy Framework: Tiered-Access}
\label{sec:policy}
This policy framework establishes a tiered system to ensure ethical, secure, and context-aware access to CoT reasoning in LLMs. By tailoring access levels and role-based authentication to distinct user groups, it balances innovation, transparency, and accountability and mitigating risks of misuse.

\paragraph{Academic Users} Academic researchers, universities, and non-commercial institutions receive full access to raw CoT outputs to support critical research initiatives, such as developing small-sized dense models or auditing LLM reasoning patterns. To prevent misuse, access is contingent on strict ethical reviews for sensitive domains, mandatory anonymization of personally identifiable information, and binding agreements prohibiting unauthorized redistribution or commercialization of CoT data. Academic users also gain access to auxiliary metadata, such as model confidence scores and uncertainty intervals, to validate the integrity of reasoning processes. This tier prioritizes open scientific inquiry while ensuring compliance with ethical and legal standards.

\paragraph{Business Users}
Enterprises and startups gain restricted CoT access through tiered licensing agreements. Licensing terms scale with risk levels, query volume, and company size, with premium pricing for high-impact use cases like customer-facing tools. Businesses are contractually barred from reverse-engineering model architectures, leveraging CoT insights for competitive espionage, or repurposing outputs to undermine industry peers. Compliance is enforced through third-party audits, dynamic API rate limits, and digital watermarking to trace leaks. Violations incur severe penalties, including fines up to 200\% of licensing fees and permanent access revocation. This tier enables commercial innovation while safeguarding against exploitation.

\paragraph{General Users} CoT outputs must provide a complete yet concise reasoning trace that logically guides the user to the conclusion, avoiding technical jargon, speculative detours, or mixed-language responses. Each trace should be structured as a step-by-step narrative (e.g., “1. Identified user intent as X; 2. Cross-referenced authoritative sources Y and Z; 3. Prioritized solution A due to alignment with evidence B”) to ensure clarity and skimmability. Crucially, every CoT output must include a bolded, upfront disclaimer explicitly stating that the model may hallucinate or generate inaccurate inferences, with a strict prohibition against relying on CoT reasoning for high-stakes domains such as medical diagnoses, financial investments, or legal decisions without verification from qualified human experts.

To enforce readability, LLM developers must implement post-processing modules that prune redundant reasoning loops, convert complex equations or code into plain-language summaries, and flag ambiguous or biased logic for revision. Regular users retain the option to toggle between simplified summaries and more granular details.  Compliance is ensured through regularly audits of CoT outputs to assess adherence to clarity and safety standards, alongside a user-driven feedback system (e.g., in-app “Report Ambiguity” buttons) to crowdsource improvements.


\paragraph{Cross-sector Implementation}
This framework requires coordinated efforts from LLM developers, regulatory bodies, and user institutions to ensure effective enforcement. LLM developers are primarily responsible for integrating technical safeguards (e.g., readability filters, bias detection algorithms) and implementing role-based authentication systems. Academic institutions must enforce ethics reviews and anonymization protocols for researchers accessing full CoT outputs, while businesses using licensed CoT data are obligated to comply with third-party audits and contractual terms prohibiting misuse. Regulatory agencies (e.g., FTC, EU AI Office) should oversee public transparency metrics and levy penalties for violations, supported by independent auditors for high-risk sectors like healthcare or finance. Finally, advisory boards—comprising ethicists, industry experts, and civil society representatives—must guide iterative policy updates, while users should report ambiguities or breaches via designated channels. Cross-sector collaboration and clear accountability structures are critical to maintaining ethical, transparent, and user-centric CoT reasoning.

\paragraph{Implementation Examples}
A startup building a medical diagnostic tool must undergo independent ethics audits, pay risk-adjusted licensing fees, and embed warnings like “This analysis is not a substitute for professional medical advice.” An academic publishing a study on CoT bias receives full reasoning traces but must redact proprietary dataset references to comply with anonymization rules.

By harmonizing accessibility, innovation, and ethical rigor, this framework fosters trust in LLMs while curbing misuse across research, public, and commercial domains.

\section{Conclusion}
This paper explores the benefits and risks of Chain-of-Thought (CoT) disclosure in LLMs and proposes a tiered-access policy framework to balance transparency, security, and accountability. By structuring CoT availability based on user categories and integrating ethical safeguards, our approach ensures responsible AI deployment.
\newpage
\appendix
\section*{Embedded Ethics Discussion}

\paragraph{Teaser for New Peers}
The ethical integration of Chain-of-Thought (CoT) reasoning in AI systems requires navigating a critical trade-off: while transparency in decision-making fosters trust and enables error correction, unrestricted access to reasoning processes risks misuse, bias propagation, and adversarial manipulation. Our tiered-access framework addresses this challenge by embedding ethical considerations directly into technical design. Researchers are granted full visibility into CoT outputs to advance model distillation and audit biases, businesses receive curated insights under contractual safeguards to prevent competitive exploitation, and general users access simplified, jargon-free explanations accompanied by disclaimers to mitigate overreliance on unverified outputs. This approach aligns with embedded ethics principles, where ethical constraints are proactively woven into system architecture rather than retrofitted as an afterthought.

\paragraph{Our Results into Pedagogy}  
\textbf{Lectures} could analyze real-world case studies, such as medical AI systems issuing flawed diagnoses due to opaque reasoning or malicious actors repurposing raw CoT outputs for misinformation campaigns, underscoring the societal stakes of ethical oversight. \textbf{Coding assignments} might task students with implementing tiered-access APIs (e.g., suppressing proprietary logic for commercial users) or developing post-processing tools to detect and remove biased or speculative steps from CoT traces, simulating the technical complexities of ethical adherence. \textbf{Projects} could further bridge theory and practice by challenging learners to design a CoT ``ethical filter'' that anonymizes sensitive data, flags hallucinated reasoning, or blocks high-stakes advice without human verification—mirroring the framework’s safeguards. \textbf{Another assignment} might involve simulating adversarial attacks on CoT outputs to expose vulnerabilities, followed by refining access policies to address these risks. These exercises emphasize that ethical AI demands collaboration among developers (designing guardrails), regulators (auditing compliance), and users (reporting misuse). By framing ethics as a dynamic, code-level concern rather than an abstract principle, the module trains future researchers and engineers to prioritize societal impact alongside technical performance, ensuring AI systems evolve as equitable tools.

\section*{Contribution Statement}
This is a joint work by Yihang Chen, Haikang Deng, Kaiqiao Han, and Qingyue Zhao (alphabetically). Yihang Chen proposes to work on the topic of the chain-of-thought disclosure policy, frames the tiered-access protocol, and writes the \cref{sec:pro} and \cref{sec:policy}. Haikang Deng surveys related papers and writes \cref{sec:motivation}. Kaiqiao Han works on the \cref{sec:con} and overall writing. Qingyue Zhao works on \cref{sec:trans,subsec:inherent}, and the Embedded Ethics Discussion.

\bibliography{iclr2025_conference}
\bibstyle{acl_natbib}

\end{document}